\documentclass{article}

\usepackage{arxiv}
\usepackage[utf8]{inputenc}
\usepackage[T1]{fontenc}
\usepackage{url}
\usepackage{booktabs}
\usepackage{amsfonts}
\usepackage{nicefrac}
\usepackage{microtype}
\usepackage{lipsum}
\usepackage{graphicx}
\usepackage{numprint}
\usepackage{titlesec}
\usepackage{XCharter}
\usepackage{amsmath}
\usepackage{adjustbox}
\usepackage{color, colortbl}
\usepackage{amssymb}
\usepackage{xcolor}
\usepackage{bbm}
\usepackage{datetime}
\usepackage[font = {footnotesize, it}]{caption}
\usepackage[ruled, vlined]{algorithm2e}
\usepackage{tabularx}
\usepackage{tablefootnote}
\usepackage{subcaption}
\usepackage{dsfont}
\usepackage{mathtools}
\usepackage{breakcites}
\usepackage{amsthm}
\usepackage{wrapfig}
\usepackage{float}
\usepackage[colorlinks = true, linkcolor = bluecite, urlcolor = bluecite]{hyperref}
\npthousandsep{,}

\newcommand{\norm}[1]{\left\lVert#1\right\rVert}

\DeclareMathOperator*{\mediane}{med}
\newcommand{\Esp}[1]{\mathbb{E}\! \left[ #1 \right]}
\newcommand{\Prr}[1]{\mathbb{P}\! \left( #1 \right)}
\newcommand{\defeq}{\vcentcolon=}
\newtheorem{Example}{Example}

\definecolor{bluecite}{HTML}{0875b7}
\hypersetup{citecolor = bluecite}

\graphicspath{ {./images/} }

\title{How much telematics information do insurers need for claim classification?}

\author{
    Francis Duval \\
    Chaire Co-operators en analyse des risques actuariels\\
    Département des mathématiques\\
    Université du Québec à Montréal\\
    Montréal, QC H2X 3Y7\\
    \texttt{duval.francis.2@courrier.uqam.ca} \\
    \And
    Jean-Philippe Boucher\\
    Chaire Co-operators en analyse des risques actuariels\\
    Département des mathématiques\\
    Université du Québec à Montréal\\
    Montréal, QC H2X 3Y7\\
    \texttt{boucher.jean-philippe@uqam.ca} \\
    \And
    Mathieu Pigeon\\
    Chaire Co-operators en analyse des risques actuariels\\
    Département des mathématiques\\
    Université du Québec à Montréal\\
    Montréal, QC H2X 3Y7\\
    \texttt{pigeon.mathieu.2@uqam.ca} \\
}

\begin{document}

\maketitle

\begin{abstract}
    It has been shown several times in the literature that telematics data collected in motor insurance help to better understand an insured's driving risk. Insurers that use this data reap several benefits, such as a better estimate of the pure premium, more segmented pricing and less adverse selection. The flip side of the coin is that collected telematics information is often sensitive and can therefore compromise policyholders' privacy. Moreover, due to their large volume, this type of data is costly to store and hard to manipulate. These factors, combined with the fact that insurance regulators tend to issue more and more recommendations regarding the collection and use of telematics data, make it important for an insurer to determine the right amount of telematics information to collect. In addition to traditional contract information such as the age and gender of the insured, we have access to a telematics dataset where information is summarized by trip. We first derive several features of interest from these trip summaries before building a claim classification model using both traditional and telematics features. By comparing a few classification algorithms, we find that logistic regression with lasso penalty is the most suitable for our problem. Using this model, we develop a method to determine how much information about policyholders' driving should be kept by an insurer. Using real data from a North American insurance company, we find that telematics data become redundant after about $3$ months or $\numprint{4000}$ kilometers of observation, at least from a claim classification perspective.
\end{abstract}


\section{Introduction}\label{sec:intro}

Usage-Based Insurance (UBI) is a type of motor insurance for which premiums are determined using information about the driving behaviour of the insured. This information is usually collected using telematics technology, most often through a device installed in the vehicle or a mobile application. Because of its multiple benefits, this type of insurance is increasingly promoted by insurers. It also seems to be more and more appreciated by consumers. A survey\footnote{\raggedright\url{www.willistowerswatson.com/en-US/Insights/2017/05/infographic-how-ready-are-consumers-for-connected-cars-and-usage-based-car-insurance}} conducted by \emph{Willis Towers Watson} on $\numprint{1005}$ insurance consumers in the United States reports that $4$ out of $5$ drivers are in favour of sharing their recent driving information in exchange for a personalized insurance product. Among the benefits, it seems clear nowadays that the addition of telematics information into the insurance pricing models improves the precision of the pure premium (see for instance \cite{ayuso2019improving}, \cite{perez2019semi}, \cite{roel2017unraveling} and \cite{lemaire2015use}). UBI also has many positive impacts on society (see for instance \cite{greenberg2009} and \cite{bordoff2008}). Indeed, because it encourages individuals to drive less and more safely, it helps making roads safer, reducing traffic congestion, limiting greenhouse gas emissions, and making insurance more affordable, among other things.

The idea of usage-based insurance was first articulated by William Vickrey, considered as the father of UBI. In \cite{vickrey1968automobile}, he criticizes the premium structure then used in motor insurance. He believes that the insurance premium should be modulated according to the use of the vehicle, and thus appear as a variable cost to the insured. In order to correct the premium structure that he considers deficient, Vickrey proposes in the late $1960$s a new type of insurance where the premium increases with usage. In particular, he suggests pricing through a tax applied to gasoline or tires: insureds who consume more gasoline (or tires) would then have a higher premium. However, due to the lack of technology and organizational barriers, it was not until the mid-$1990$s that the first UBI program appeared in the United States. Nowadays, several major general insurance companies have their own UBI program, and this type of insurance is still growing in popularity. It is now a fact that UBI is collectively beneficial in many ways, and that is why it seems to be the future of motor insurance.

A fairly general section of the UBI literature discusses the feasibility, implemention, costs and benefits of UBI. \cite{litman2007} explores the practicability, pros and cons of differents types of distance-based insurance, such as \textit{Milage Rate Factor} and \textit{Pay-at-the-Pump}. \cite{greenberg2009} establish that every mile driven insured with UBI rather than conventional insurance brings a significant benefit to the community, which they quantify at \$$0.016$. With this in mind, he proposes benefit-maximizing rules and incentives to increase the size of the UBI market. \cite{bordoff2008} estimate that the change from traditional insurance to UBI reduces mileage by $8$\% in average, resulting in yearly savings of \$$658$ per vehicle. Indeed, the fact that the premium increases with usage is a strong incentive to drive less. They state that most of these savings are attributable to reduced congestion and accidents. 

In the recent literature on automobile insurance pricing, among all information you can get from a telematics device, several papers only focus on the distance driven, which is probably the most useful measure for ratemaking. \cite{boucher2017exposure} simultaneously analyze the impact of distance driven and duration on claim frequency using Generalized Additive Models (GAMs) for cross-sectional count data. They find that neither distance nor duration is proportional to frequency, but that frequency tends rather to stabilize once a certain distance or duration has been reached. The authors invoke a ``learning effect'' and a ``highway effect'' to explain the fact that a policyholder travelling $2x$ miles is less than twice as dangerous as an insured travelling $x$ miles. \cite{boucher2020longitudinal} go further and analyse the link between distance driven and frequency using models for panel count data, including a Generalized Additive Model for Location, Scale and Shape (GAMLSS) and a GAM with fixed effects. They refute the idea of the ``learning effect'' to explain the non-linearity of frequency. Instead, they find that the relationship between frequency and distance driven is approximately linear, and that the apparent non-linearity is due to residual heterogeneity incorrectly captured by GAMs.

Since an insured rarely has more than one claim per year (see for instance \cite{boucher2007risk}\footnote{Table 2 of this paper informs us that 99.5\% of the policyholders have 1 claim or less.}), assigning a probability of claiming is almost like assigning a premium. Therefore, a significant amount of studies related to UBI ratemaking have focused on claim classification. \cite{pesantez2019predicting} compare the predictive performance of a non-penalized logistic regression and a boosting algorithm called XGBoost using classical and telematics information, including distance driven, fraction of driving at night, fraction of driving in urban areas and fraction of driving above speed limit. In \cite{huang2019automobile}, the authors compare the performance of various classification algorithms while proposing a way to bin continuous telematics variables in order to create a finite number of risk classes and thus increase interpretability. It turns out that this discretization also increases the predictive power of the algorithms. \cite{paefgen2013evaluation} also compare multiple classification algorithms on real claim classification data and propose a novel way to aggregate telematics information into what they call an ``aggregate risk factor''.

Moreover, the question of how to transform telematics data into useful information for pricing is still a thorny issue to this day. Indeed, it is not yet clear how to use raw telematics information in an optimal way. This falls within the field of feature engineering. Some simply create features manually from the raw data, while others contributed using techniques from machine learning, deep learning and pattern recognition (\cite{weidner2016classification}, \cite{gao2018feature}, \cite{gao2018driving}, \cite{gao2019convolutional}).

Despite its many benefits, the growing popularity of UBI means that insurers are accumulating large amounts of sensitive data about their insureds. \cite{dewri2013inferring} have shown that it is possible to deduce users' destinations by analyzing their driving data, without even having access to the geographical coordinates of the trips. In recent years, concerns have been arising about the use of telematics technology, particularly with respect to confidentiality and data usage. In Canada and in other countries, insurance regulators make recommendations regarding UBI products. In particular, Financial Services Commission of Ontario (FSCO) and Financial Services Regulatory Authority of Ontario (FSRA) state that telematics data fall under the definition of ``personal data'', and must therefore be handled according to the relevant legislation (see \cite{howell2013}). They also mention that the insurer must inform the consumer in several respects, such as the type of information collected and the use made of it. In spite of these recommendations, the fact remains that large amounts of personal data are stored by insurance companies. From an ethical, practical and economical point of view, it is nowadays important for an insurer to keep a minimal amount of personal data on its insureds. Notably, collecting less data reduces storage and manipulation costs in addition to facilitate data leakage prevention.

This raises the question of how much information an insurer should collect on the driving behavior of its insureds, which should be just enough to get a good idea of how the insured drives, but not too much for the reasons cited in the the preceding paragraph. We explore this avenue through the binary claim classification framework, where the goal is to assign each insured a probability of claiming. In addition to data traditionally used in motor insurance pricing, we have access to telematics data in a format where we have one observation per trip, which allows us to derive interesting and representative features of the insured's driving. We first build a classification model using both classical and telematics features. To this effect, we determine which classification algorithm is the most suited to our problem by comparing among a logistic regression with lasso penalty, a logistic regression with elastic-net penalty and a random forest, all this while accounting for interactions between features. It turns out that the lasso model is best suited to our task. Using this classification model, we then develop a method for determining how much telematics information an insurer should collect. For this purpose, we derive several classification datasets using increasing amounts of information about the insured's driving. We then fit a lasso model on each of these datasets and compare performance: telematics information is considered redundant when it no longer substantially improves the classification score.

In Section~\ref{sec:data}, we describe the classical and telematics data available to us in conjunction with details of how telematics features are derived. Section~\ref{sec:preprocessing} follows, where we show how the classification datasets are built using varying amounts of information as well as other preprocessing steps. Then, in Section~\ref{sec:classification_algorhtms}, the mathematical framework of supervised classification is introduced, in addition to explaining the functioning of the $3$ preselected algorithms, namely the two penalized logistic regressions and the random forest. This is followed by Section~\ref{sec:analyzes}, in which we first make a choice among the classification models presented in Section~\ref{sec:classification_algorhtms}. We choose the lasso logistic regression for its good performance and for its simplicity. We also find that adding interactions does not substantially improve the performance of the models. Using our classification model as well as the datasets built in Section~\ref{sec:preprocessing}, we then develop a method to determine the right amount of telematics information to collect based on non-parametric bootstrapping. Using the data provided by the North American insurer, we find that telematics information no longer improves substantially classification after about $3$ months, or $\numprint{4000}$ kilometers of observation. Finally, we conclude in Section~\ref{sec:conclusion}.

\section{Data}\label{sec:data}

\begin{figure}[ht]
    \centering
    \includegraphics[width = 0.9\textwidth]{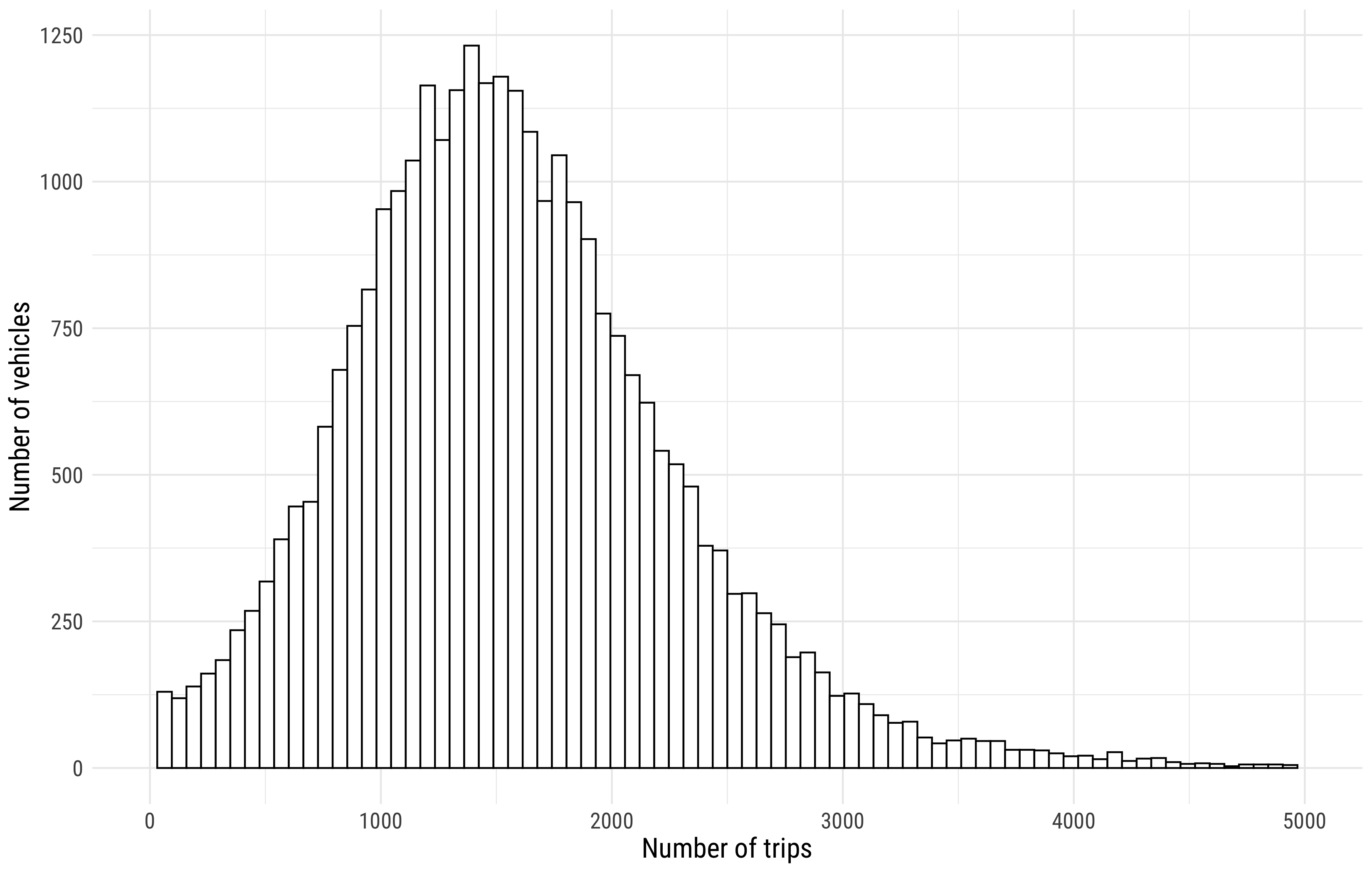}
    \caption{Histogram of the number of trips for the $\numprint{29799}$ vehicles in their observed year.}
    \label{fig:hist_nb_trips}
\end{figure}

All the data is provided by a North American insurer and is related to personal auto insurance in the province of Ontario. We have access to a telematics database consisting of $\numprint{210854092}$ summaries of trips made by $\numprint{67355}$ vehicles between January 2016 and December 2018, for which an extract is shown in Table~\ref{tab:1}.
\begin{table}[ht]
    \centering
    \begin{adjustbox}{max width = \textwidth}
        \begin{tabular}{l c c c c c}
            \toprule 
            \textbf{VIN} & \textbf{Trip number} & \textbf{Departure datetime} & \textbf{Arrival datetime} & \textbf{Distance} & \textbf{Maximum speed}\\ 
            \midrule
            A & $1$ & $2016$-$04$-$09$ {$15$:$23$:$55$} & $2016$-$04$-$09$ {$15$:$40$:$05$} & $10.0$ & $72$\\
            A & $2$ & $2016$-$04$-$09$ {$17$:$49$:$33$} & $2016$-$04$-$09$ {$17$:$57$:$44$} & $4.5$ & $68$\\
            \vdots & \vdots & \vdots & \vdots & \vdots & \vdots \\
            A & $3312$ & $2019$-$02$-$11$ {$18$:$33$:$07$} & $2019$-$02$-$11$ {$18$:$54$:$10$} & $9.6$ & $65$\\
            \cmidrule(l){1-6}
            B & $1$ & $2016$-$04$-$04$ {$06$:$54$:$00$} & $2016$-$04$-$04$ {$07$:$11$:$37$} & $14.0$ & $112$\\
            B & $2$ & $2016$-$04$-$04$ {$15$:$20$:$19$} & $2016$-$04$-$04$ {$15$:$34$:$38$} & $13.5$ & $124$\\
            \vdots & \vdots & \vdots & \vdots & \vdots & \vdots \\
            B & $2505$ & $2019$-$02$-$11$ {$17$:$46$:$47$} & $2019$-$02$-$11$ {$18$:$19$:$22$} & $39.0$ & $130$\\
            \cmidrule(l){1-6}
            C & $1$ & $2016$-$01$-$16$ {$15$:$41$:$59$} & $2016$-$01$-$16$ {$15$:$51$:$35$} & $3.3$ & $65$ \\
            \vdots & \vdots & \vdots & \vdots & \vdots & \vdots \\
            \bottomrule 
        \end{tabular}
    \end{adjustbox}
    \caption{Extract from the telematics dataset. Dates are displayed in the yyyy-mm-dd format.} 
    \label{tab:1} 
\end{table}
The recording of a trip, made using an On-Board Diagnostics (OBD) device, begins when the vehicle is turned on and stops when the ignition is turned off. Each trip is summarized in $4$ measurements: the datetime of departure and arrival, the distance driven and the maximum speed reached. Trips are also associated with a vehicle via the vehicle identification number (VIN), but there is no column to identify the insured person, which makes it impossible to know who the driver is. Therefore, our analysis is based on vehicles rather than policyholders. For $\numprint{57671}$ of these vehicles, which are all observed during one or more insurance contracts, we have access to features traditionally used in motor insurance (gender, age, region, etc.) as well as claiming information. Among these features, that we will call ``classical features'', $10$ were selected and are described in Table~\ref{tab:2}. Distributions of the classical numeric features and those of the classical categorical features are shown in Figures~\ref{fig:dist_num} and \ref{fig:dist_cat} of Appendix~\ref{appendix}, respectively.
\begin{table}[ht]
    \centering
    \begin{adjustbox}{max width = \textwidth}
        \begin{tabular}{l l l}
            \toprule 
            \textbf{Classical feature name} & \textbf{Description} & \textbf{Type} \\
            \midrule
            \texttt{annual\_distance} & Annual distance declared by the insured & Numeric\\
            \texttt{commute\_distance} & Distance to the place of work declared by the insured person & Numeric\\
            \texttt{conv\_count\_3\_yrs\_minor} & Number of minor contraventions in the last 3 years & Numeric\\
            \texttt{gender} & Gender of the policyholder & Categorical\\
            \texttt{marital\_status} & Marital status of the policyholder & Categorical\\
            \texttt{pmt\_plan} & Payment plan chosen by the policyholder & Categorical\\
            \texttt{veh\_age} & Vehicle age & Numeric\\
            \texttt{veh\_use} & Use of the vehicle & Categorical\\
            \texttt{years\_claim\_free} & Number of years since last claim & Numeric\\
            \texttt{years\_licensed} & Number of years since driving licence & Numeric\\
            \bottomrule 
        \end{tabular}
    \end{adjustbox}
    \caption{Classical features selected for the analysis.} 
    \label{tab:2} 
\end{table}
Only vehicles having at least one full-year contract are kept for the analysis, which means we end up with $\numprint{29799}$ vehicles. In the following section, the telematics dataset will be aggregated by contract (which means the latter will go from one row per trip to one row per contract) and then merged with classical features and claim information (which are already on a contract basis). The resulting dataset will then be given as input to supervised learning algorithms. Since we know that supervised learning algorithms generally learn best on independent observations, we keep only the earliest one-year contract for each vehicle, which allows us to get rid of the dependency that exists between the different contracts associated with the same vehicle. In their observed year, $99.8$\% of the vehicles have made less than $\numprint{5000}$ trips, for an average of $\numprint{1581}$ trips per vehicle. The complete distribution is shown in Figure~\ref{fig:hist_nb_trips}.\\

\begin{figure}[ht]
    \centering
    \includegraphics[width = \textwidth]{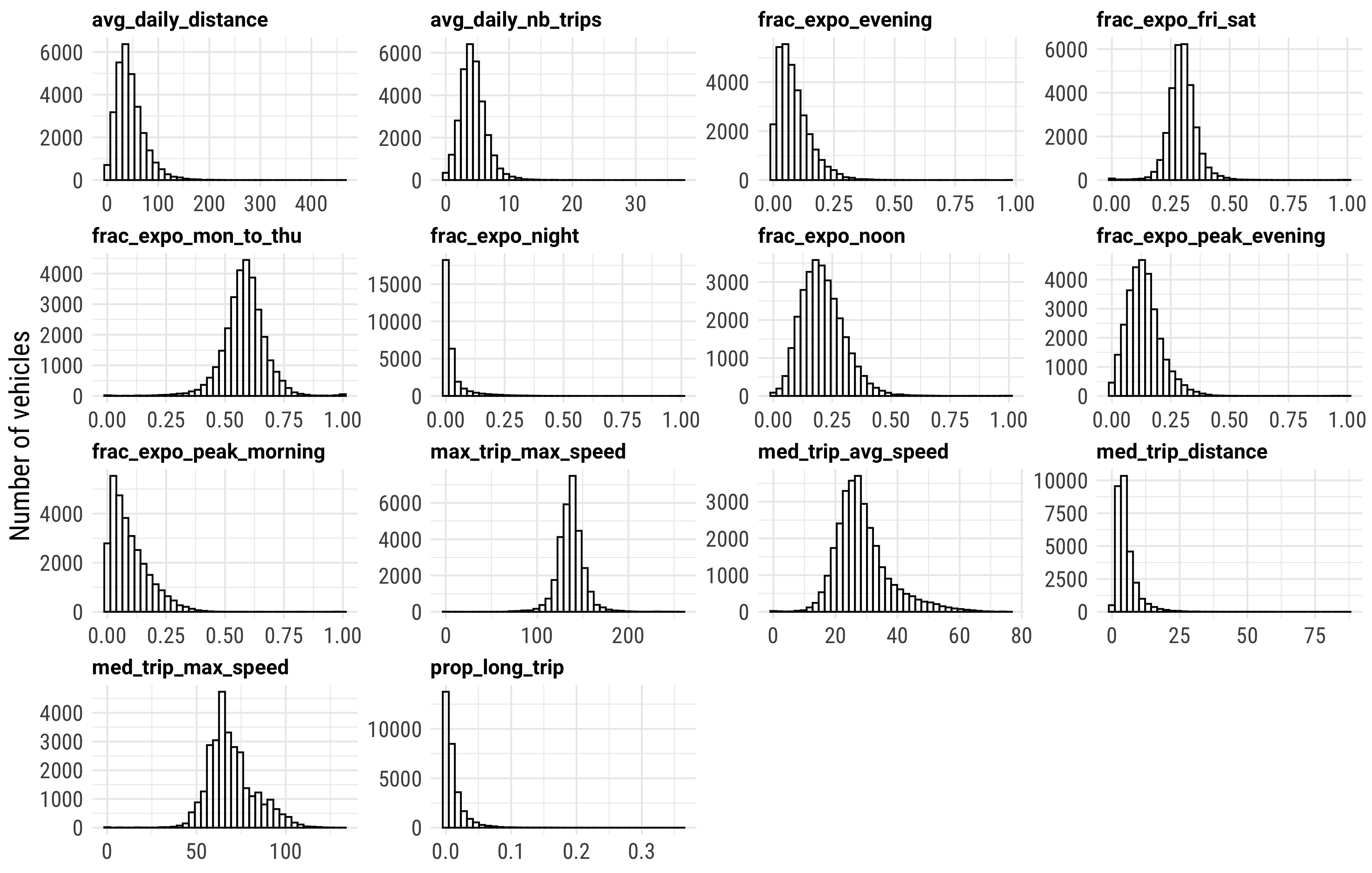}
    \caption{Distributions over the $\numprint{29799}$ vehicles of the 14 features extracted from the telematics dataset.}
    \label{fig:hist}
\end{figure}

Based on the trip summaries of Table~\ref{tab:1}, we wish to derive telematics features that best depict the insured's driving behavior by aggregating the trips for each vehicle. Indeed, we want these features, or covariates, to have a good predictive power when inputted into a supervised classification algorithm. This falls within the field of feature extraction, which is often a crucial step in machine learning. However, extracting (or creating) features from raw telematics data in an optimal way is not a simple task and is a research avenue in itself. This problem is addressed in several articles such as \cite{wuthrich2017covariate} and \cite{gao2018feature}. Features extracted in these two studies nevertheless require second-by-second data, which we do not have at hand. We are therefore largely inspired by features of the ``usage'', ``travel habits'' and ``driving performance'' types derived in \cite{huang2019automobile}. From the telematics dataset, we thus extract a total of $14$ features, all described in Table~\ref{tab:3} and for which the distribution is shown in Figure~\ref{fig:hist}. 
\begin{table}[ht]
    \centering
    \begin{adjustbox}{max width = \textwidth}
        \begin{tabular}{l l c c r}
            \toprule 
            & & \multicolumn{2}{c}{\textbf{Mean value}} &\\
            \cmidrule(l){3-4}
            \textbf{Feature name} & \textbf{Description} & Non-claimants ($94.7$\%) & Claimants ($5.3$\%) & \textbf{p-value (t-test)}\\ 
            \midrule
            \textbf{avg\_daily\_distance} & \textbf{Average daily distance} & $\boldsymbol{44.3}$ & $\boldsymbol{50.1}$ & $\boldsymbol{<0.0001}$\\
            \textbf{avg\_daily\_nb\_trips} & \textbf{Average daily number of trips} & $\boldsymbol{4.3}$ & $\boldsymbol{4.8}$ & $\boldsymbol{<0.0001}$\\
            \textbf{med\_trip\_avg\_speed} & \textbf{Median of the average speeds of the trips} & $\boldsymbol{28.8}$ & $\boldsymbol{28.0}$ & $\boldsymbol{<0.0001}$\\ 
            med\_trip\_distance & Median of the distances of the trips & $5.2$ & $5.2$ & $0.9203$\\
            med\_trip\_max\_speed & Median of the maximum speeds of the trips & $69.6$ & $70.0$ & $0.2906$\\ 
            \textbf{max\_trip\_max\_speed} & \textbf{Maximum of the maximum speed of the trips} & $\boldsymbol{138.2}$ & $\boldsymbol{141.9}$ & $\boldsymbol{<0.0001}$\\ 
            prop\_long\_trip & Proportion of long trips ($>100\text{km}$) & $0.0108$ & $0.0103$ & $0.4114$\\ 
            frac\_expo\_night & Fraction of night driving\tablefootnote{0h-6h} & $0.0262$ & $0.0276$ & $0.1630$\\ 
            \textbf{frac\_expo\_noon} & \textbf{Fraction of midday driving}\tablefootnote{11h-14h} & $\boldsymbol{0.210}$ & $\boldsymbol{0.198}$ & $\boldsymbol{<0.0001}$\\ 
            \textbf{frac\_expo\_evening} & \textbf{Fraction of evening driving}\tablefootnote{20h-0h} & $\boldsymbol{0.0845}$ & $\boldsymbol{0.0965}$ & $\boldsymbol{<0.0001}$\\ 
            frac\_expo\_peak\_morning & Fraction of morning rush hour driving\tablefootnote{7h-9h Monday to Friday} & $0.0968$ & $0.0985$ & $0.4453$\\
            \textbf{frac\_expo\_peak\_evening} & \textbf{Fraction of evening rush hour driving}\tablefootnote{17h-20h Monday to Friday} & $\boldsymbol{0.137}$ & $\boldsymbol{0.143}$ & $\boldsymbol{<0.0001}$\\ 
            frac\_expo\_mon\_to\_thu & Fraction of driving on Monday to Thursday & $0.582$ & $0.582$ & $0.6786$\\
            frac\_expo\_fri\_sat & Fraction of driving on Friday and Saturday & $0.299$ & $0.300$ & $0.5475$\\
            \bottomrule 
        \end{tabular}
    \end{adjustbox}
    \caption{Mean value of the $14$ features extracted from the telematics dataset for claimants and non-claimant. Two-sample t-tests were conducted to determine whether the mean differs significantly between the two groups.} 
    \label{tab:3} 
\end{table}
In Table~\ref{tab:3}, we also display the average value of the telematics features for two groups of vehicles, namely the claimants (those who have claimed at least once during their observed year) and the non-claimants (those who have not claimed during their observed year), and a two-sample t-test is conducted for each of the features to determine whether the mean differs significantly between the two groups. It turns out that the difference in the mean is significant (at a $95$\% confidence level) for half of the $14$ features, which we have highlighted in bold in Table~\ref{tab:3}. This suggests that these $7$ features contain predictively relevant information. Note that claimant vehicles tend to travel more distance and make more trips, which seems natural. They also tend to have a lower average median speed. This may be due to the fact that claimant insureds have a higher propensity to drive in the city, where average speed is lower and the risk of collision, higher than elsewhere. Claimants also tend to have a higher maximum speed reached in their observed year, drive less in the midday and more in the evening, especially in the rush hour. Note that we have at our disposal $\numprint{2}$ ``mileage'' variables. The first one, \texttt{annual\_distance}, is declared by the insured at the very beginning of the contract and corresponds to the number of kilometers that the insured estimates he/she will drive during the following year. The second, \texttt{avg\_daily\_distance}, is measured by the OBD device and is the actual number of kilometers the insured has driven during the year, divided by $365.25$, which the average number of days in a year. One could argue that these two features tell roughly the same thing about an insured and that we should get rid of one of them. However, insureds underestimate their distance travelled in a year by an average $\numprint{1399}$ kilometers. Indeed, the average of the \texttt{avg\_daily\_distance} feature times $365.25$ is $\numprint{16374}$ kilometers while the average of \texttt{annual\_distance} is $\numprint{14975}$ kilometers. Moreover, the Pearson correlation coefficient between these two is only $0.37$, which means there is a considerable discrepancy between \texttt{annual\_distance} and \texttt{avg\_daily\_distance}. We therefore keep both of these features in our analysis, since each one tells a different story about an insured. In order to determine which features (classical and telematics combined) are significant in predicting the occurrence of a claim, a non-penalized logistic regression was fitted to the data. The coefficients obtained for the 24 features and their standard deviations are shown in Figure~\ref{fig:glm_coefs} of Appendix~\ref{appendix}. It turns out that $10$ of the $24$ coefficients, among which $7$ are related to telematics features, are significant (using a 5\% threshold).
\begin{Example}
	In order to illustrate the exact calculation of the $14$ telematics features, let us imagine that a insured, during a given contract that we assume lasts $7$ days, made only $5$ trips, summarized in Table~\ref{tab:ass_fictif}.
    \begin{table}[ht]
        \centering
        \begin{adjustbox}{max width = \textwidth}
            \begin{tabular}{c c c c c c c}
                \toprule 
                \textbf{Trip number} & \textbf{Departure time} & \textbf{Arrival time} & \textbf{Weekday} & \textbf{Distance} & \textbf{Average speed} & \textbf{Maximum speed}\\ 
                \midrule
                $1$ & {$18$:$20$}& {$18$:$28$} & Monday & $8$ & $60$ & $73$\\
                $2$ & {$17$:$40$}& {$17$:$54$} & Monday & $9$ & $39$ & $70$\\
                $3$ & {$09$:$35$}& {$09$:$48$} & Tuesday & $17$ & $78$ & $102$\\
                $4$ & {$07$:$30$}& {$07$:$37$} & Thursday & $9$ & $77$ & $92$\\
                $5$ & {$12$:$20$}& {$13$:$35$} & Saturday & $109$ & $88$ & $120$\\
                \bottomrule 
            \end{tabular}
        \end{adjustbox}
        \caption{Summaries of the $5$ trips made by a fictitious insured. Note that weekday and average speed can be easily derived from the telematics dataset.} 
        \label{tab:ass_fictif} 
    \end{table}
    The calculation of the telematics features related to this insured would then be done according to Table~\ref{tab:14}.
    \begin{table}[ht]
        \centering \scriptsize
        \begin{adjustbox}{width = 0.8\textwidth}
            \begin{tabular}{l r c}
                \toprule 
                \textbf{Feature} & \textbf{Computation} & \textbf{Value}\\
                \midrule
                \texttt{avg\_daily\_distance} & $(8 + 9 + 17 + 9 + 109)/7$ & $21.7$\\ 
                \texttt{avg\_daily\_nb\_trips} & $5/7$ & $0.71$\\
                \texttt{med\_trip\_avg\_speed} & $\mediane\{60, 39, 78, 77, 88\}$ & $77$\\
                \texttt{med\_trip\_distance} & $\mediane\{8, 9, 17, 9, 109\}$ & $9$\\
                \texttt{med\_trip\_max\_speed} & $\mediane\{73, 70, 102, 92, 120\}$ & $92$\\ 
                \texttt{max\_trip\_max\_speed} & $\max\{73, 70, 102, 92, 120\}$ & $120$\\ 
                \texttt{prop\_long\_trip} ($>100\text{km}$) & $1/5$ & $0.2$\\ 
                \texttt{frac\_expo\_night} & $0/(8 + 9 + 17 + 9 + 109)$ & $0$\\ 
                \texttt{frac\_expo\_noon} & $109/(8 + 9 + 17 + 9 + 109)$ & $0.72$\\ 
                \texttt{frac\_expo\_evening} & $0/(8 + 9 + 17 + 9 + 109)$ & $0$\\ 
                \texttt{frac\_expo\_peak\_morning} & $9/(8 + 9 + 17 + 9 + 109)$ & $0.06$\\
                \texttt{frac\_expo\_peak\_evening} & $(9 + 8)/(8 + 9 + 17 + 9 + 109)$ & $0.11$\\ 
                \texttt{frac\_expo\_mon\_to\_thu} & $(8 + 9 + 17 + 9)/(8 + 9 + 17 + 9 + 109)$ & $0.28$\\
                \texttt{frac\_expo\_fri\_sat} & $109/(8 + 9 + 17 + 9 + 109)$ & $0.72$\\
                \bottomrule 
            \end{tabular}
        \end{adjustbox}
        \caption{Telematics features calculated for the fictitious insured of Table~\ref{tab:ass_fictif}.} 
        \label{tab:14} 
    \end{table}
    \normalsize
\end{Example}

One must be careful when analyzing telematics data. Indeed, insureds who have chosen to be observed telematically for insurance purposes do not correspond to the general population of insureds, which means our data cannot be considered a simple random sample of the company's insured population. As far as we are concerned, 10\% to 15\% of the insureds in our North American insurance company's portfolio are observed with a telematics device, and these are generally worse and younger drivers. This is because at the time the data was collected, in order to encourage policyholders to choose the telematics option, insurers were offering a 5\% entry discount in addition to a renewal discount ranging from 0\% to 25\%. At that time, it was also not possible for insurers, at least in the region where the data was collected, to increase the premium based on telematics information: the latter could only be reduced. Because car insurance in Ontario is very expensive, any discount is welcomed by high-premium policyholders such as bad drivers and youths. As a consequence, an unexpectedly large proportion of bad/young drivers end up using the telematics option. However, this selection bias does not affect our analysis since the models and methods we develop apply only to telematically observed insureds. One must only be careful not to draw conclusions from these data and apply them to the general population of insureds.

\section{Data preparation}\label{sec:preprocessing}

\subsection{Design of the classification datasets}

With the data we have at hand, we wish to build several classification datasets using a varying amount of telematics information, or trip summaries. We will later on perform classification on each of them and compare performance, which will allow us to determine how much telematics information is needed to obtain a proper classifier. For this purpose, we compute telematics features of Table~\ref{tab:3} in several versions, which we do in two different ways. The first way to proceed, called the ``time leap'' method (TL), consists in first calculating features using only one month of trip summaries for each vehicle, and then add one month worth of data to each subsequent version. Since vehicles are observed over one year, telematics features of Table~\ref{tab:3} are derived in $12$ different versions. In general, the $k^\text{th}$ version is calculated using the first $k$ months of telematics information related to a given vehicle, for $k = 1, \dots, 12$. The second way to proceed, called the ``distance leap'' method (DL), is quite similar to the time leap way, but uses $\numprint{1000}$-kilometer leaps instead of one-month leaps to jump from version to version. For the sake of uniformity, we also derive telematics features in $12$ different versions using this second method. In general, the $k^\text{th}$ version of a telematics feature for vehicle $i$ is calculated using the first $\numprint{1000}\times k$ kilometers of telematics information related to this vehicle, for $k = 1, \dots, 12$. If it turns out that the vehicle has done less than $\numprint{1000}k$ kilometers in its observed year, we simply use all available telematics information from this vehicle. Note that $37$\% of the vehicles have accumulated less than $\numprint{12000}$ kilometers of driving during their observed year, which means that they end up with some identical versions of the telematics features for the distance leap method. For instance, a vehicle that has accumulated $\numprint{5500}$ kilometers of driving has its distance leap versions $6$ to $12$ of the telematics fatures calculated with the same amount of telematics information, namely with $\numprint{5500}$ kilometers of trip summaries. Each version of the telematics features can be represented by a $\numprint{29799} \times \numprint{14}$ matrix, where each row corresponds to a vehicle and each column to a feature. The matrix containing the $k^\text{th}$ version of the telematics features derived according to the time leap method is noted $\boldsymbol{x}_k^\text{TL}$, while the one derived according to the distance leap method is noted $\boldsymbol{x}_k^\text{DL}$. Let us also denote by $\boldsymbol{x}^c$ the $\numprint{29799} \times \numprint{10}$ matrix containing the classical features of Table~\ref{tab:2} and by $\boldsymbol{y}$ the response vector of length $\numprint{29799}$, which indicates whether or not each vehicle had a claim in its observed year. By using time leap versions of the telematics features, we then build $12$ classification datasets, denoted $\mathcal{D}_1^\text{TL}, \dots, \mathcal{D}_{12}^\text{TL}$, all sharing the same classical features $\boldsymbol{x}^c$ as well as the same response vector $\boldsymbol{y}$. The only difference between them is the version of the telematics features used or, in other words, the amount of trip summaries used to compute telematics features. In general, $\mathcal{D}_k^\text{TL}$, $k = 1, \dots, 12$, is the classification dataset built with the matrix of telematics features $\boldsymbol{x}_k^\text{TL}$, which means it is obtained by concatenating $\boldsymbol{x}^c$, $\boldsymbol{x}_k^\text{TL}$ and $\boldsymbol{y}$. In addition to these $12$ classification datasets, we also create a dataset containing no telematics information, noted $D_0^\text{TL}$. The latter is therefore built by concatenating $\boldsymbol{x}^c$ and $\boldsymbol{y}$. Note that all $13$ datasets describe the same vehicles and therefore have the same number of rows. In a similar fashion, $13$ datasets $\mathcal{D}_0^\text{DL}, \dots, \mathcal{D}_{12}^\text{DL}$ are built using the distance leap versions of the telematics features. In order to properly test models, it is common in machine learning to split the observations into training and testing sets. For future use, we thus randomly draw $70$\% of the $\numprint{29799}$ vehicles to make up the training set, and the remaining $30$\% forms the testing set. Training and testing datasets are respectively denoted by $\mathcal{T}_k^m$ and $\mathcal{V}_k^m$, where $k = 0, \dots, 12$ and $m \in \{\text{TL}, \text{DL}\}$.

\subsection{Preprocessing}

\begin{wrapfigure}{R}{0.6\textwidth}
    \begin{center}
        \includegraphics[width=0.5\textwidth]{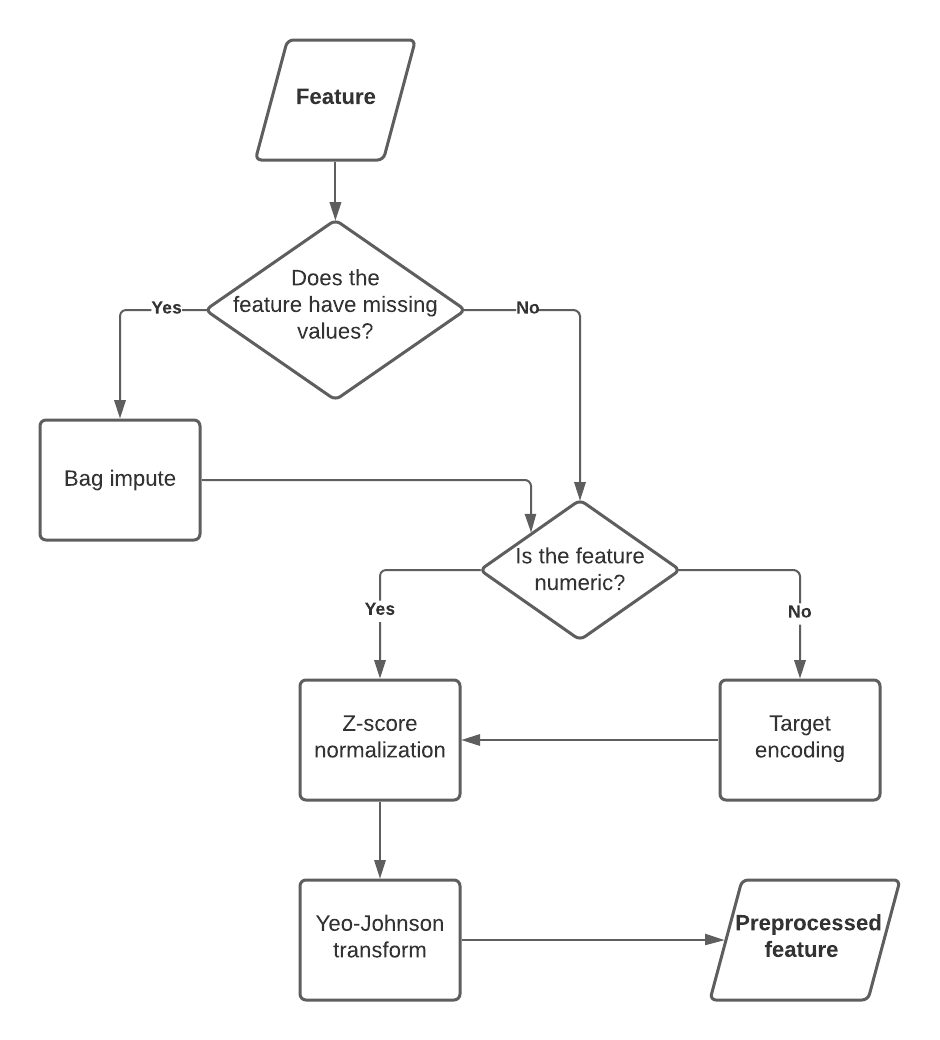}
    \end{center}
    \caption{Flowchart of the feature preprocessing ``recipe'', that is applied every time a model is trained or validated in this analysis.}
    \label{fig:preprocessing_flow_chart}
\end{wrapfigure}

Training and testing datasets need to be preprocessed every time they are being fed to the models, either for training or scoring purposes. The reasons for this are manifold. First, some of our features are categorical (\texttt{gender}, \texttt{marital\_status}, \texttt{pmt\_plan} and \texttt{veh\_use}), and many supervised learning algorithms cannot handle this type of information, which means we need a way to encode them numerically. For this, we choose an embeding method called ``target encoding'', which consists in replacing the value of each category, that is originally a string of characters, by a real number based on the response (or target) column. In the special case of mean target encoding, the value of each observation is replaced by the mean of the response variable for the category to which that observation belongs. For instance, imagine we have the feature ``gender'' with categories ``woman'' and ``man'' and that the mean of the response variable is $0.09$ for women and $0.11$ for men. Women would then be encoded with the value $0.09$ while men would be assigned the value $0.11$. What we use is similar to mean encoding, except that the encoded values are derived using a GLM rather than using the mean. Suppose we are in the context of supervised binary classification and we want to encode the feature $x$, which is categorical with $k$ categories. GLM target encoding consists in first fitting a logistic regression without intercept using only $x$ to explain the binary response variable $y$, which yields coefficient estimates $\widehat{\beta}_1, \dots, \widehat{\beta}_k$. Then, each of the $k$ categories is encoded with its corresponding coefficient. Hence, category $j$ is encoded with the value $\widehat{\beta_j}$, for $j = 1, ..., k$. This means that the $k$-category categorical feature $x$ becomes a numerical feature with $k$ unique values. In order to perform target encoding, the \texttt{step\_lencode\_glm} function of the \texttt{embed} package in the \texttt{R} programming language is used. Note that prior target encoding the categorical features, rare categories (i.e., those associated with $5$\% of the observations or less) are lumped together in a catch-all category. 

Secondly, the \texttt{commute\_distance} feature, which is numerical, is missing for $22.3$\% of the observations. Since many classification algorithms cannot handle missing values, we need a way to impute them. Virtually any prediction algorithm could be used to perform imputation, but some are more suitable than others. We choose an algorithm based on bagged decision trees (see \cite{breiman1996bagging}) as they are known to be good at imputing missing data, partly because they generally have good accuracy and because they do not extrapolate beyond the bounds of the training data. Bagged decision trees are also preferable to random forest because they require fitting fewer decision trees to have a stable model. The idea is to first consider the feature to be imputed, namely \texttt{commute\_distance}, as a response variable. Then, a bagged decision tree model is trained on all observations for which \texttt{commute\_distance} is not missing, using all features except the latter as predictors. This fitted model is then scored on all observations with a missing \texttt{commute\_distance} value, and the prediction is used as a replacement value. To implement bag imputation, we use the \texttt{step\_bagimpute} function of the \texttt{recipe} package in \texttt{R}. 

Thirdly, supervised learning algorithms generally learn best when the data is preprocessed in a certain way. For instance, they generally benefit from normalized and rather symmetrical feature distributions. To normalize features, we use the z-score normalization, which is a quite popular choice. Consider the numerical feature vector $\boldsymbol{x} = (x_1, \dots, x_n)$. The z-score normalized version of this vector is 
\begin{align*}
    \boldsymbol{x}^* = \left(\frac{x_1 - \overline{x}}{s}, \dots, \frac{x_n - \overline{x}}{s}\right),
\end{align*}
where $\overline{x} = \frac{1}{n}\sum_{i = 1}^n x_i$ and $s = \sqrt{\frac{1}{n - 1}\sum_{i = 1}^n (x_i - \overline{x})^2}$ are the empirical mean and standard deviation of vector $\boldsymbol{x}$, respectively. In order to obtain more symmetric feature distributions, we use the Yeo-Johnson transformation (see \cite{yeo2000new}), which is similar to the Box-Cox transformation except that it allows for negative input values. Yeo-Johnson transformation also has the effect of making the data more normal distribution-like. The Yeo-Johnson transformation $\psi$ applied on a real value $x$ is defined as follows:
\begin{align*}
    \psi(x, \lambda) =
    \begin{cases}
      ((x + 1)^\lambda - 1)/\lambda & \text{if } \lambda \ne 0, x \ge 0\\
      \ln(x + 1) & \text{if } \lambda = 0, x \ge 0\\
      -[(-x + 1)^{2 - \lambda} - 1]/(2 - \lambda) & \text{if } \lambda \ne 2, x < 0\\
      -\ln(-x + 1) & \text{if } \lambda = 2, x < 0, \\
    \end{cases}
\end{align*}
where $\lambda$ is a parameter that is optimized by maximum likelihood so that the resulting empirical distribution resembles a normal distribution as closely as possible. Note that the preprocessing steps are performed in a specific order and depends on the feature type. Figure~\ref{fig:preprocessing_flow_chart} illustrates the data preprocessing ``recipe''.

\section{Binary classification algorithms}\label{sec:classification_algorhtms}

\subsection{Binary classification framework and classification algorithm preselection}

Let us be in the context of binary supervised classification, in which we have at our disposal $n$ labeled examples $\{(\boldsymbol{x}_i, y_i)\}_{i = 1}^n$, where $\boldsymbol{x}_i = (x_{i1}, \dots, x_{ip})$ and $y_i \in \{0, 1\}$ are respectively the $p$-dimensional vector containing the features and the label (or response variable) for observation $i$. The goal is to estimate $\Esp{Y|\boldsymbol{x}}$, the conditional mean of $Y$ given the features, which can also be seen as the conditional probability $\Prr{Y = 1|\boldsymbol{x}}$. Many algorithms have been developed to estimate this probability, including logistic regression, random forest, artificial neural networks, support vector machines, etc. For the analysis, we retain two supervised learning algorithms, i.e. penalized logistic regression and random forest. The reason for this choice is that the latter often give excellent classification results while requiring little preprocessing of the data. In fact, these two could be qualified as ``off-the-shelf'' algorithms because they perform implicitly feature selection and do not require much data preprocessing, unlike other algorithms such as neural networks. They are also quite easy to tune because they do not have too many hyperparameters. Note that some algorithms that require more care before inputting the data into the model can probably lead to better classification performance, but our goal is not really in that way. For the penalized logistic regression, we consider two specifications, namely the lasso penalty (also called $\ell_1$-penalty) and the elastic-net penalty.

\subsection{Logistic regression}

In logistic regression, the goal is to approximate the conditional probability of having a positive case ($Y = 1$) by applying the sigmoid function over a linear transformation of the features. The model can therefore be expressed as
\begin{align}
    p_i \defeq \Prr{Y_i = 1|\boldsymbol{x}_i} = \sigma\left(\beta_0 + \sum_{j = 1}^p \beta_j x_{ij}\right),
    \label{eq:predictionlasso}
\end{align}
where $\sigma$, the sigmoid function, ensures that the output is a real number between 0 and 1. The model is parametrized by the unknown parameter vector $\boldsymbol{\beta} = (\beta_1, \dots, \beta_p)$, and the intercept $\beta_0$. These parameters are often estimated by maximum likelihood, which leads to asymptotically unbiased estimates. Maximizing likelihood is equivalent to minimizing cross-entropy loss, given by
\begin{align}
    L(\beta_0, \boldsymbol{\beta}) &= -\frac{1}{n} \sum_{i = 1}^n y_i \ln(p_i) + (1 - y_i) \ln(1 - p_i).
    \label{eq:crossentropyloss}
\end{align}
There is no closed formula for maximum likelihood parameter estimates in the logistic regression framework, but a variety of numerical optimization methods can be used. Most of the time, the method of iteratively reweighted least squares (IRLS) is used. 

\subsection{Lasso logistic regression}

Maximum likelihood estimator is the asymptotically unbiased estimator with the smallest variance. However, it is rarely the best for prediction accuracy. Indeed, although it has a low bias, it has a rather large variance. In $1996$, \cite{tibshirani1996regression} proposes a new method called \textit{least absolute shrinkage and selection operator} (lasso) for estimating parameters in linear regression that reduces the variance of the parameters at the cost of increased bias. In practice, this decrease in variance more than offsets the increase in bias, thus improving predictive performance. Although this method was originally used for models using the least squares estimator, it generalizes quite naturally to generalized linear models. In the case of logistic regression, a penalty proportional to the sum of the absolute values of the parameters is added to the cross-entropy loss of Equation~\ref{eq:crossentropyloss}. In lasso logistic regression, the optimization problem thus becomes
\begin{align}
    \min_{\beta_0, \boldsymbol{\beta}} \left\{L(\beta_0, \boldsymbol{\beta}) + \lambda \sum_{j = 1}^p |\beta_j|\right\},
    \label{eq:logisticlassolagrangian}
\end{align}
where $\lambda \ge 0$ is the penalty hyperparameter. In the special case where $\lambda = 0$, the penalty term disappears and we recover the conventional non-penalized logistic regression. This penalty hyperparameter is not directly optimized by the algorithm and must therefore be chosen by the user, for instance with cross-validation and grid search. Equation~\ref{eq:logisticlassolagrangian} is called the Lagrangian formulation of the lasso logistic regression optimization problem. It can be useful to rewrite this in the constrained form,
\begin{align}
    \min_{\beta_0, \boldsymbol{\beta}} \left\{L(\beta_0, \boldsymbol{\beta})\right\} \quad\text{s.t.}\quad \sum_{j = 1}^p |\beta_j| \le s.
    \label{eq:logisticlassoconstrained}
\end{align}
Note that there is a one-to-one correspondance between $\lambda$ and $s$. This formulation allows to realize that the model gives itself a ``budget'' of parameters. Indeed, the sum of the absolute values of the coefficients, which is the $\ell_1$ norm of the parameter vector $\boldsymbol{\beta}$, must always be less than or equal to the constant $s$ set by the user. This has the effect of shrinking and even zeroing some of the logistic regression coefficients. The lasso logistic regression fits into the more general framework where the constraint on the parameter vector is given by
\begin{align}
    \sum_{j = 1}^p |\beta_j|^q \le s,
\end{align}
where $q \ge 0$ is a fixed real number. In particular, setting $q = 1$ retrieves the lasso constraint, whereas $q = 0$ and $q = 2$ correspond respectively to best subset selection and Ridge logistic regressions. Best subset selection and Rigde have both their pros and cons. Ridge regression only shrinks coefficients: it never sets them to zero, which makes interpretation more difficult. In general, one prefers to have a sparse model: according to Occam's Razor Principle, a simple explanation of a phenomenon is to be preferred to a more complex one. Best subset selection leads to sparse models, but it involves resolving a nonconvex and combinatorial optimization problem, and becomes infeasible above about $50$ features. Lasso regression attempts to retain good features from both subset selection and Ridge: it leads to sparse models while being a convex optimization problem. In fact, we can show that $q = 1$ is the smallest value that leads to a convex problem, and this partly explains why lasso regression is so popular. The loss function to be minimized in Equation~\ref{eq:logisticlassolagrangian} is not differentiable due to the absolute values in the penalty term, but it is convex, and a wide range of methods from convex optimzation theory have been developped to compute the solution, including coordinate descent, subgradient and proximal gradient-based methods. In this paper, lasso logistic regression is fit using the \texttt{glmnet} library of the \texttt{R} programming language. This library uses a proximal Newton algorithm, which consists in making a quadratic approximation of the log-likelihood part of the loss function and then applying a weighted coordinate descent, iteratively. For more details about lasso logistic regression, we refer to \cite{friedman2010regularization}, \cite{hastie2015statistical} and \cite{hastie2016introduction}.

\subsection{Elastic-net logistic regression}

Even though lasso regression often performs very well on tabular data, it has a few drawbacks. Among other things, when there is a group of features that are highly correlated with each other, the lasso tends to select only one feature in the group and does not care which one is selected. This can be a problem for us, as some of the telematics features we have created (or even classical features) may be highly correlated with each other (e.g. \texttt{avg\_daily\_distance} and \texttt{avg\_daily\_nb\_trips}). \cite{zou2005regularization} address this problem by proposing a new regularization and variable selection method called ``elastic-net''. The elastic-net regression combines the penalties of Ridge and lasso regressions and thereby retains the best of both worlds. Ridge regression is known to share parameters among highly correlated features, which often improves performance, while lasso yields sparse models and thus performs feature selection, which is desirable. Elastic-net regression thus yields sparse models while improving the treatment of highly correlated features. More precisely, elastic-net regression includes a penalty term in its loss function that is a linear combination of the $\ell_1$ (lasso) and $\ell_2$ (Ridge) penalties. In the particular case of binary classification, elastic-net coefficients are therefore obtained by solving
\begin{align}
    \min_{\beta_0, \boldsymbol{\beta}} \left\{L(\beta_0, \boldsymbol{\beta}) + \lambda \left[\frac{(1 - \alpha)}{2} \norm{\boldsymbol{\beta}}_2^2 + \alpha \norm{\boldsymbol{\beta}}_1 \right]\right\},
    \label{eq:elasticnetloss}
\end{align}
where a new ``mixing'' hyperparameter $0 \le \alpha \le 1$ appears. Ridge and lasso regressions are in fact special cases of elastic-net regression, when $\alpha = 0$ and $\alpha = 1$, respectively. If $\alpha$ and $\lambda$ are known, Criterion~\ref{eq:elasticnetloss} is convex and can be solved by a variety of algorithms such as coordinate descent.

\subsection{Random forest}

Random forest classifier was first formalized by Leo Breiman in \cite{breiman2001random} and enjoy great popularity for its many strengths, including a usually high accuracy on structured data. It consist in building several decision trees on slightly modified versions of the original dataset. The final prediction is then obtained by aggregating all the trees, often by taking the mean on the individual predictions. The trees built are usually very deep and therefore have little bias, but have a large variance. Aggregating them allows to drastically decrease the variance and to obtain a much more flexible prediction function, increasing the predictive power compared to a single tree. The main advantage of random forest over logistic regression is that it can approximate a wider range of functions since it is a non-parametric algorithm, making fewer assumptions about the form of the underlying data generating function. Another benefit is that it automatically takes into account the interactions between features due to the tree structure of its components. Like lasso and elastic-net logistic regressions, random forest has an embeded feature selection mechanism. On the other hand, random forest is harder to interpret, so for equal performance, logistic regression is preferred. Note that since logistic regression assumes a linear relationship between the features and the log-odd of a positive case, random forest usually outperforms it when this relationship is rather non-linear. In order to train a random forest model, one first generates bootstrap samples of the training dataset on which decision trees will later be built. This is done by drawing observations with replacement, and usually, as many observations as there are in the original training set are drawn. Then, for each of these bootstrap samples, $1 \le p^* \le p$ features are randomly picked, $p^*$ being a previously chosen hyperparameter. This last step allows the decision trees to be built on different subsets of features, which has the effect of decorrelating the predictions, thus improving performance. Finally, a CART-like classification tree (see \cite{breiman1984classification}) is built on each of these datasets, each one yielding an estimated probability of having a positive case for every point of the feature space. The criterion we use to build the trees is the impurity of the nodes, measured by the Gini index. Every time the feature space is split in two, we thus choose the splitting point that decreases the Gini index the most. Other criteria are also possible. For a given point, a final prediction is obtained by averaging the individual predictions of all trees. More details about the general procedure are given in Algorithm~\ref{alg:1}. For more information about random forest, we refer to \cite{breiman2001random}, \cite{hastie2015statistical} and \cite{hastie2016introduction}.
\begin{algorithm}[ht]
    \SetAlgoLined    
    
    \SetKwInOut{Input}{Inputs}
    \SetKwInOut{Output}{Output}

    \Input{
        \begin{itemize}
            \item[$\bullet$] Training dataset $\mathcal{T} = \{(\boldsymbol{x}_i, y_i)\}_{i = 1}^n$ containing $p$ features
            \item[$\bullet$] Number of features to pick $1 \le p^* \le p$
            \item[$\bullet$] Number of trees (or bootstrap samples) $B$ 
        \end{itemize}
    }
    
    \For{$b = 1, \dots, B$}{
        \begin{enumerate}
            \item Generate a bootstrap sample $\mathcal{T}^*$ by drawing with replacement $n$\\ observations from $\mathcal{T}$.
            \item Pick at random $p^*$ of the $p$ features.
            \item Build a CART classification tree on $\mathcal{T}^*$ using only the $p^*$ features previously picked,\\ yielding the prediction function $\widehat{T}_b(\boldsymbol{x})$.
        \end{enumerate}
    }
    
    \Output{Random forest classifier $\hat{f}_\text{RF}(\boldsymbol{x}) = \frac{1}{B} \sum_{b = 1}^B \widehat{T}_b(\boldsymbol{x})$}

 \caption{Random forest binary classifier}
 \label{alg:1}
\end{algorithm}

\section{Analyzes}\label{sec:analyzes}

\subsection{Claim classification model}

In order to choose among the $3$ classification models presented in Section~\ref{sec:classification_algorhtms} (lasso logistic regression, elastic-net logistic regression and random forest), we make them compete on the dataset whose telematics features are computed with all available trip summaries of the observed year for each vehicle, namely $\mathcal{D}_{12}^\text{TL}$. 

\subsubsection{Hyperparameter tuning}\label{hyperparametertuning}

First of all, the $3$ models need to be tuned since they all involve hyperparameters that are not directly optimized by their respective algorithm. The general idea behind tuning is to test several combinations of hyperparameters and evaluate the out-of-sample performance for each of them, which is often done using a validation set or cross-validation on the training set. One then usually chooses the combination of hyperparameters yielding the best performance. We choose to use $5$-fold cross-validation with the Area Under the receiver operating characteristic Curve (AUC) as the metric for evaluating classification performance. This metric is often used in binary classification, notably because it does not depend on the threshold used for classification and because it works well on unbalanced datasets (our dataset is highly unbalanced since there are many more non-claimant vehicles than claimant ones). Different methods have been developed to choose which combinations of hyperparameters to try out, including grid search, random search, Bayesian optimization, gradient-based optimization and evolutionary optimization. For penalized logistic regression, it is generally not necessary to use a sophisticated tuning algorithm, and one usually proceeds with a simple grid search, which we do. For the random forest, we choose a more refined method, namely a Bayesian optimization method, who have been shown to yield better results than grid search and random search (see for instance \cite{snoek2012practical}). 

The lasso requires the tuning of only one hyperparameter, which is the penalty parameter $\lambda$ of Equation~\ref{eq:logisticlassolagrangian}. We first create a grid of $100$ penalty values ranging from $10^{-10}$ to $1$ uniformly distributed on a logarithmic scale, namely the set
\begin{align*}
    \mathcal{G}_\lambda = \left\{10^{-10 + \frac{i - 1}{9.9}}\right\}_{i = 1}^{100}.
\end{align*}
Then, $5$-fold cross-validation AUC is assessed on the training set $\mathcal{T}_{12}^\text{TL}$ using each of the values in $\mathcal{G}_\lambda$ as a candidate. It turns out that the best value for $\lambda$ is $0.000231$ (which is the value in $\mathcal{G}_\lambda$ associated with $i = 64$), yielding an AUC of $0.6373$. For the elastic-net model, the mixing parameter $\alpha$ must also be tuned in addition to the penalty parameter (see Equation~\ref{eq:elasticnetloss}). With a grid search, one usually uses a coarse uniform grid of values for $\alpha$. We thus choose to use $5$ values uniformly distributed between $0$ and $1$ inclusively, namely the grid $\mathcal{G}_\alpha = \{0, 0.25, 0.5, 0.75, 1\}$. For $\lambda$, we use the same grid as for lasso, i.e. $\mathcal{G}_\lambda$. The performance of the $|\mathcal{G}_\lambda| \times |\mathcal{G}_\alpha| = 100 \times 5 = 500$ possible combinations of hyperparameters is thereafter evaluated and it turns out that $\lambda = 0.00298$ (which is the value in $\mathcal{G}_\lambda$ associated with $i = 75$) and $\alpha = 0$ is the best choice, with an AUC value of $0.6377$, slightly better than lasso. 

Regarding the random forest, two hyperparameters are tuned, namely the number of features drawn every time a tree is built $p^*$ (see Algorithm~\ref{alg:1}) and the minimum number of observations required to make a further split in any leaf $n^*$. Note that for simplicity, the latter does not appear in Algorithm~\ref{alg:1}. The total number of trees $B$ must also be chosen, but it is not strictly speaking a hyperparameter. It only needs to be large enough for the performance to stabilize. We choose $B = 1000$, which is plenty for the number of observations we have. Note that $B$ cannot be too large since a random forest can never overfit due to too many trees. However, increasing the number of trees obviously increases the computation time. Bayesian optimization is used to find the best possible pair $(p^*, n^*)$. Basically, it consists in treating the unknown function that maps hyperparameter values to the loss function evaluated on a test set (or with cross-validation) as a random function. An \textit{a priori} distribution, which captures beliefs about the behavior of this function, is defined. Then, as combinations of hyperparameters are tested and evaluated, the \textit{a priori} distribution is updated, yielding the \textit{a posteriori} distribution. The latter is thereafter used to find the next combination of hyperparameters to try out. So unlike grid and random search, Bayesian optimization leverages past evaluations to find the most promising candidates faster. To implement this procedure, we use the \texttt{tune\_bayes} function of the \texttt{tune} package in \texttt{R}, which uses a Gaussian process to model the probability distribution over the function. One can think of the Gaussian process as a generalization of the normal distribution concept to functions. It turns out that $(p^*, n^*) = (1, 39)$ is the best pair that has been tested, yielding an AUC value of $0.6004$. This means only one feature is picked every time a tree is built and that the growth of a tree stops when all its leaves have less than $39$ observations. 

\subsubsection{Interactions}

A limitation of logistic regression is that it does not naturally take into account interactions between features, unlike random forest. Fortunately, this can be overcome by manually calculating interactions. According to the interaction hierarchy principle (see \cite{kuhn2019feature}), lower-level interactions are more likely than higher-level ones to explain variation in the response variable. For instance, level $2$ interactions are more likely to be predictive than level $3$ interactions, which are more likely to be predictive than level $4$ interactions, and so on. Therefore, in order to keep computation time reasonable, we only consider level $2$ (or pairwise) interactions. We also limit ourselves to calculating the interactions between the $10$ classical features of Table~\ref{tab:2} as well as telematics features whose mean value is significantly different between claimant and non-claimant vehicles (i.e. the $7$ bolded features in Table~\ref{tab:3}). These $7$ features are presumed to have good predictive power, and according to the principle of heredity (see \cite{kuhn2019feature}), they have a higher probability than other features of creating interactions that also have good predictive power. This entails calculating $\binom{17}{2} = 136$ pairwise interactions. Next, lasso and elastic-net regressions are tuned on the training dataset $\mathcal{T}_{12}^\text{TL}$ expanded with the $136$ interactions as new columns. For this purpose, the same grid search procedure described in Section~\ref{hyperparametertuning} is used.

\subsubsection{Out-of-sample performance comparison}

Optimal hyperparameter(s) found as well as the $5$-fold cross-validation AUC value for each of the $5$ tuned models are shown in Table~\ref{tab:results_classification}. The interaction-free elastic-net model has the best cross-validation score, with an AUC of $0.6377$. However, it does not outperform the lasso model, which has an AUC of $0.6373$, enough to justify the extra complexity. Indeed, an elastic-net model takes more time to tune since it has an additional hyperparameter, and also takes longer to fit. Note also that with or without interactions, the two logistic regressions perform similarly considering the variability of the AUC. Since one always prefer the simplest model for equal performance, we reject the two models including interactions. Finally, the random forest is the worst model, with an AUC of only $0.6004$, which is considerably lower than the penalized logistic regressions. We therefore reject the latter. Note that the Bayesian optimization algorithm has found that the optimal value for the hyperparameter $p^*$ is $1$, reinforcing the belief that interactions between features do not carry useful information for classification. Indeed, the fact that $p^* = 1$ means that the decision trees that make up the random forest are built with only one feature at a time, thus eliminating the possibility of including interactions.

Ideally, in order to properly estimate performance in supervised learning, a model should be evaluated on samples that have not yet been used to build or fine-tune it. We therefore use the testing dataset $\mathcal{V}_{12}^\text{TL}$ to assess the performance of the $5$ tuned models. The models are first trained on the full training dataset $\mathcal{T}_{12}^\text{TL}$ before being scored on $\mathcal{V}_{12}^\text{TL}$, which allows us to compute an AUC value for each of them, shown in Table~\ref{tab:results_classification}.
\begin{table}[ht]
    \centering
    \begin{adjustbox}{width = \textwidth}
        \begin{tabular}{l c c c c c c}
            \toprule 
            & \multicolumn{4}{c}{\textbf{Optimal hyperparameters}} & &\\
            \cmidrule(l){2-5}
            \textbf{Models} & $\lambda$ & $\alpha$ & $p^*$ & $n^*$ & \textbf{AUC (5-fold cross-validation)} & \textbf{AUC (testing set)}\\
            \midrule
            Lasso & $2.31 \times 10^{-4}$ & -- & -- & -- & $0.6373^{(0.0052)}$ & $0.6189$\\
            Elastic-net & $2.98 \times 10^{-3}$ & $0$ & -- & -- & $0.6377^{(0.0049)}$ & $0.6176$\\
            Random forest & -- & -- & $1$ & $39$ & $0.6004^{(0.0064)}$ & $0.5889$\\
            Lasso (with interactions) & $1.18 \times 10^{-3}$ & -- & -- & -- & $0.6350^{(0.0050)}$ & $0.6214$\\
            Elastic-net (with interactions) & $1.52 \times 10^{-2}$ & $0$ & -- & -- & $0.6359^{(0.0046)}$ & $0.6198$\\
            \bottomrule 
        \end{tabular}
    \end{adjustbox}
    \caption{Tuning results on the training set $\mathcal{T}_{12}^\text{TL}$ and classification performance on the testing set $\mathcal{V}_{12}^\text{TL}$. Numbers in superscript indicate standard deviations.} 
    \label{tab:results_classification} 
\end{table}
The AUC values on the testing set are slightly lower than those obtained by cross-validation, which is normal. In fact, the relatively close values indicate that we did not leak too much information into the models. The lasso model with interactions have the best testing set AUC value ($0.6214$), but we believe it is not enough to justify the addition of the $136$ interaction columns to process. From now on, we will consequently use the lasso logistic regression model without interactions. Note that the AUC values obtained are around $0.6$, which is in concordance with the literature on claim classification (see for instance \cite{huang2019automobile} and \cite{baecke2017value}).

\subsubsection{Feature importance}

\begin{figure}[ht]
    \centering
    \includegraphics[width=\textwidth]{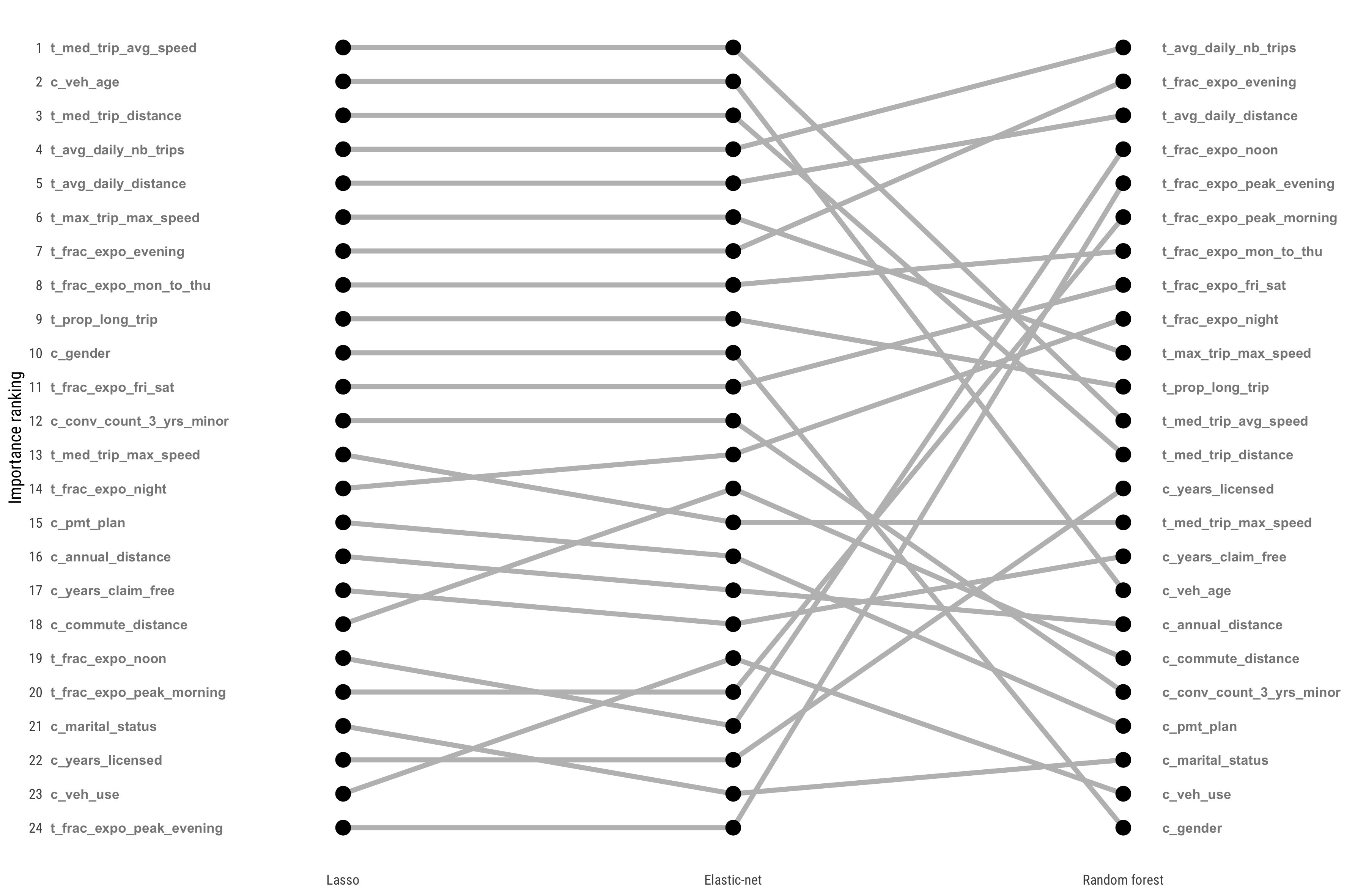}
    \caption{Ranking of the features according to their importance for each of the $3$ models. Telematics features have been given the prefix ``t\_'', while classical features have the prefix ``c\_''.}
    \label{fig:features_importance_ranking}
\end{figure}

Once the models are trained, in addition to the performance, it is interesting to look at which features contributed the most to classify observations. For the $3$ models compared, it is possible to calculate an importance score for each feature. For lasso and elastic-net logistic regressions, since the models are trained with normalized versions of the features, the absolute value of the estimated parameter may be used for this purpose. For instance, if the estimated parameter associated with \texttt{avg\_daily\_distance} is $\widehat{\beta}_1$, its importance score is $|\widehat{\beta}_1|$. For the random forest, there are many ways to compute feature importance. We choose to use the mean decrease of the Gini index. This method consists in assessing for each feature how much it has contributed to decrease the impurity of the tree nodes, measured with the Gini index. Once the importance score is obtained for all the features, we can order them from the most to the least important, which we did in Figure~\ref{fig:features_importance_ranking} for each of the $3$ models.
Looking at this figure, we can notice that the lasso and the elastic-net models have a high degree of agreement since they consider the same $12$ most important features and that the links intersect very little. This is not so surprising as these two logistic regression models work in a similar way. On the other hand, the random forest has a lower degree of agreement with the latter two since the links intersect a lot. This is probably because a random forest makes no assumptions about the nature of the relationship between the features and the response variable, whereas logistic regression assumes that this relationship is logistic-linear. The random forest therefore probably leverages better the features with rather non-linear relationships with the response variable, in contrast to logistic regression. However, all $3$ models agree that \texttt{avg\_daily\_nb\_trips}, \texttt{avg\_daily\_distance}, \texttt{max\_trip\_max\_speed}, \texttt{frac\_expo\_evening} and \texttt{frac\_expo\_mon\_to\_thu} are important features for prediction, all ranked in the top $10$ most important features. One last important thing to note is that telematics features are considered more important by the models than their classical counterparts. Indeed, telematics features have an average ranking of $9.45$, while for classical features, this value is $16.77$. This makes us believe that telematics data tells a better story about an insured's risk than traditional ratemaking information, which is consistent with the literature. For instance, \cite{roel2017unraveling} show using GAMs that telematics features have better predictive power than their classical counterparts. Indeed, their GAM model which only uses telematics outperforms their GAM using only classical factors, at least according to some criteria. In traditional pricing, i.e. in pricing without driving data, gender is an important factor in determining an insured's premium. Note that here, gender only ranks $10^\text{th}$, and this is probably due to the fact that gender is a proxy for other factors related to driving habits. As pointed out by \cite{ayuso2016telematics}, it seems that the difference in the risk of accident between men and women is mainly due to the difference in driving intensity, which is captured by two of our telematics features, namely \texttt{avg\_daily\_nb\_trips} and \texttt{avg\_daily\_distance}.

\subsection{Classification performance assessment on the 13 classification datasets}

Remember that our main objective is to develop a method to estimate the amount of telematics information that an insurer should collect from its policyholders. The method we propose consists in first tuning and training a lasso model on each of the training datasets $\mathcal{T}_0^m, \dots, \mathcal{T}_{12}^m$, $m \in \{\text{TL}, \text{DL}\}$ derived in Section~\ref{sec:preprocessing}. The lasso model tuned and trained on dataset $\mathcal{T}_k^m$ is denoted by $\mathcal{M}_k^m$, for $k = 0, \dots, 12$ and $m \in \{\text{TL}, \text{DL}\}$. We then assess the performance of these fitted models on their corresponding testing dataset. For instance, the fitted model $\mathcal{M}_k^m$ is assessed on the testing datset $\mathcal{V}_k^m$, for $k = 0, \dots, 12$ and $m \in \{\text{TL}, \text{DL}\}$. The performance is evaluated using the AUC and in order to obtain a distribution of the latter, a non-parametric bootstrap strategy is used. Non-parametric bootstrap is a method used to estimate the distribution of any statistic and consists in generating new samples (or datasets) called ``bootstrap samples''. A bootstrap sample is simply obtained by drawing with replacement as many observations as there are in the original sample. An empirical distribution is then obtained by calculating the desired statistic on each bootstrap sample. Therefore, we generate $b = 500$ bootstrap samples for each of the $13$ testing dataset $\mathcal{V}_0^m, \dots, \mathcal{V}_{12}^m$, $m \in \{\text{TL}, \text{DL}\}$. Then, in order to obtain an AUC distribution for model $\mathcal{M}_k^m$, we score the latter on each of the $500$ bootstrap samples related to $\mathcal{V}_k^m$, noted $\{^{(j)}\mathcal{V}_k^m\}_{j = 1}^{500}$ and we derive the $500$ corresponding AUC values, which form the empirical distribution. Once the empirical distribution of the AUC has been obtained for each of the models $\mathcal{M}_0^m, \dots, \mathcal{M}_{12}^m$, $m \in \{\text{TL}, \text{DL}\}$, it is thereafter possible to inspect them and determine at what point telematics information becomes redundant or, in other words, at what point the addition of telematics information in the lasso model no longer meaningfully improve the classification performance.
\begin{figure}[ht!]
    \centering
    \includegraphics[width=0.8\textwidth]{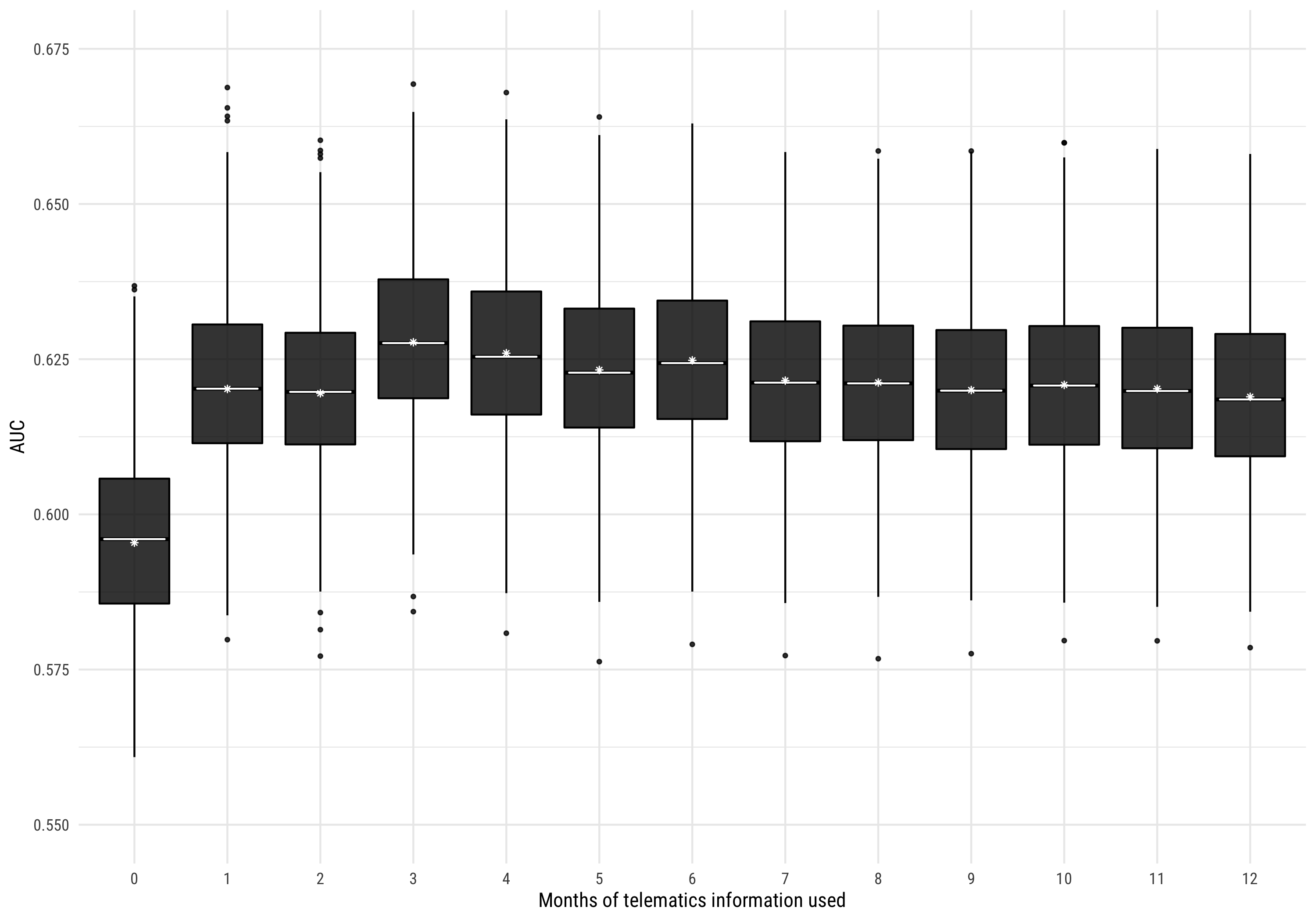}
    \hfill
    \includegraphics[width=0.8\textwidth]{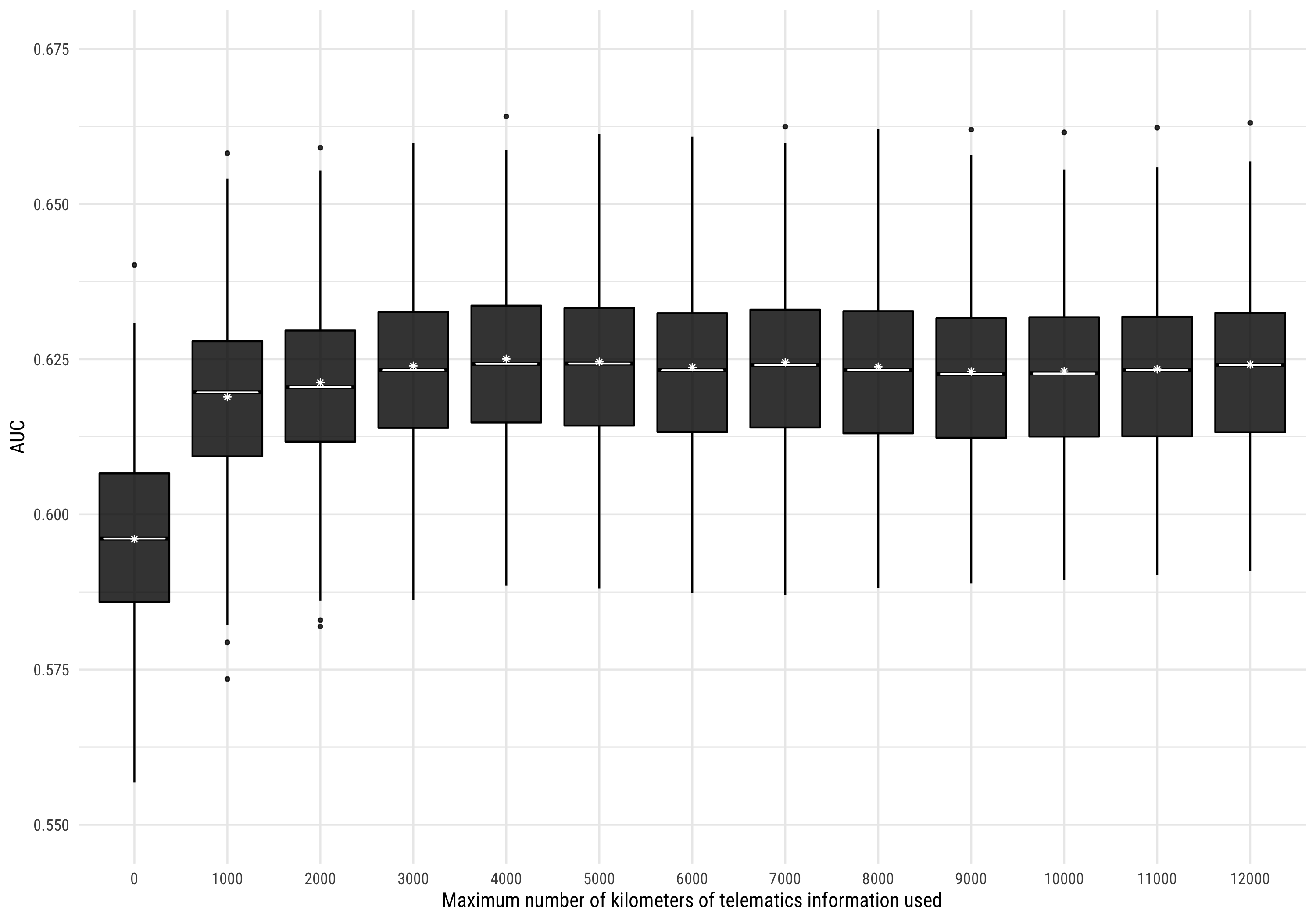}
    \caption{Distribution of the classification performance (AUC) obtained with non-parametric bootstrap for the $13$ models and for both approaches. White stars show the AUC value obtained on the original testing set.}
    \label{fig:bootstrap_results}
\end{figure}
The AUC distributions are shown in Figure~\ref{fig:bootstrap_results} for the $13$ lasso models and for both approaches, namely the time leap approach (upper panel) and the distance leap approach (bottom panel). The first thing that can be noticed when looking at the boxplots is that the addition of telematics features into the model significantly improves its performance, which is in line with the literature. Indeed, the first boxplot from the left in each of the upper and bottom panels, which corresponds to the classical model, is lower than all the other ones. Looking at the upper panel, we realize that with the addition of as little as one month's worth of trip summaries, we can drastically improve classification performance. Similarly, adding just $\numprint{1000}$ kilometers of telematics data into the classical model also improves performance substantially (bottom panel). After $1$ month or $\numprint{1000}$ kilometers, although the marginal improvement is less substantial, the AUC slightly increases but stabilizes fairly quickly, after about $3$ months or $\numprint{4000}$ kilometers. After this point, telematics information collected seems redundant and no longer improves classification. This suggests that telematics features calculated with $3$ months of trip summaries have about the same predictive power as those calculated with $1$ complete year of data.

\section{Conclusions}\label{sec:conclusion}

In this paper, we first created several relevant telematics features from trip summaries in order to incorporate telematics information into supervised classification models. Then, using these features in conjunction with features traditionally used in automobile insurance, and by adequately preprocessing the data, we compared the performance of $3$ popular classification algorithms, namely a lasso logistic regression, an elastic-net logistic regression and a random forest. We found that random forest, which often gives good results in classification tasks, performs the worst, while the two logistic regressions are on equal footing. However, we chose the lasso as our classification model because of its greater simplicity. We also considered interactions between features, and found that they contain little or no predictive power since their addition into the models does not improve out-of-sample classification performance. Then, based on the lasso model and thus remaining within the framework of supervised classification, we developed a novel method for determining when information on the insured's driving becomes redundant. A great strength of our method is that it requires little computational time and does not require second-by-second trip data, which are large and therefore time-consuming to process and difficult to manipulate. Also, it can be used by any insurance company that has access to a dataset similar to the one used in the analysis, namely a telematics dataset where each observation corresponds to a trip. Using real data from a North American insurance company, we found out, using non-parametric bootstraping, that after about $3$ months or $\numprint{4000}$ kilometers of observation, telematics no longer help achieving a better classification performance, at least if measured with the AUC. This means it is probably not worth for this insurer to observe its policyholders during long periods of time. Rather, it is better off observing them during a predetermined short period. Indeed, people generally do not enjoy being tracked, and telematics data is both costly and risky to store and manipulate. In addition, collecting less telematics information can help to meet the recommendations of insurance regulators and facilitate the prevention of data leakage.

As reported in \cite{bolderdijk2011effects}, policyholders tend to adopt better driving habits in the early months of telematics observation when a financial incentive is given. Therefore, it is probably not a good idea to observe an insured's driving habits for 3 months (or \numprint{4000} kilometers), determine a discount (or surcharge) and then apply that same discount (or surcharge) \textit{ad vitam aeternam}. This could indeed result in pure premiums that are too low, since the observed behavior would be biased due to a lack of financial incentive to drive well after the observation period. A simple solution to this issue would be to monitor policyholders at random times during their contract, so as to eventually collect 3 months (or 4000 kilometers) of data. \cite{guillen2021near} address this issue by making pricing dynamic. Basically, they propose a pricing scheme in which the premium is updated weekly, which drives the insured to adopt good driving habits on an ongoing basis. They combine a baseline premium related to traditional risk factors with extra charges related to driving behaviors deemed unsafe captured by what they call ``near-miss events'', such as hard braking and smartphone use when driving. Moreover, they provide an alternative way to solve the problem of minimizing data storage since their method only requires keeping one week of telematics data. However, unlike our paper, they do not use information related to driving habits such as the proportion of night driving, average speed, proportion of long trips, etc., which have a demonstrated link to claim experience. Their method also requires collecting good near-miss data, which is not trivial. In this analysis, only collision coverage claims were considered, i.e. the target column given as input to the clasification models is the indicator of a collision claim, at-fault or not. One could repeat the analysis by considering at-fault and non at-fault collision claims separately, and see whether either type needs more of less telematics history to have a good estimate of the claiming probability. The analysis could also be performed again using collision-free claims, including theft, vandalism and fire, and see if we come to similar conclusions. Lastly, we could generalize the approach for count data. Therefore, instead of having the indicator of a claim as the response variable, we would instead have the number of claims, moving us into a counting regression context.

\newpage

\section*{Acknowledgement}

The authors gratefully acknowledge The Co-operators for both financial support and for providing the data used in this paper through the Co-operators Chair in Actuarial Risk Analysis. The authors also thank the Natural Sciences and Engineering Research Council of Canada for funding.


\bibliographystyle{apalike}  
\bibliography{references}


\newpage
\appendix

\section{Appendix}\label{appendix}

\begin{figure}[ht]
    \centering
    \includegraphics[width = \textwidth]{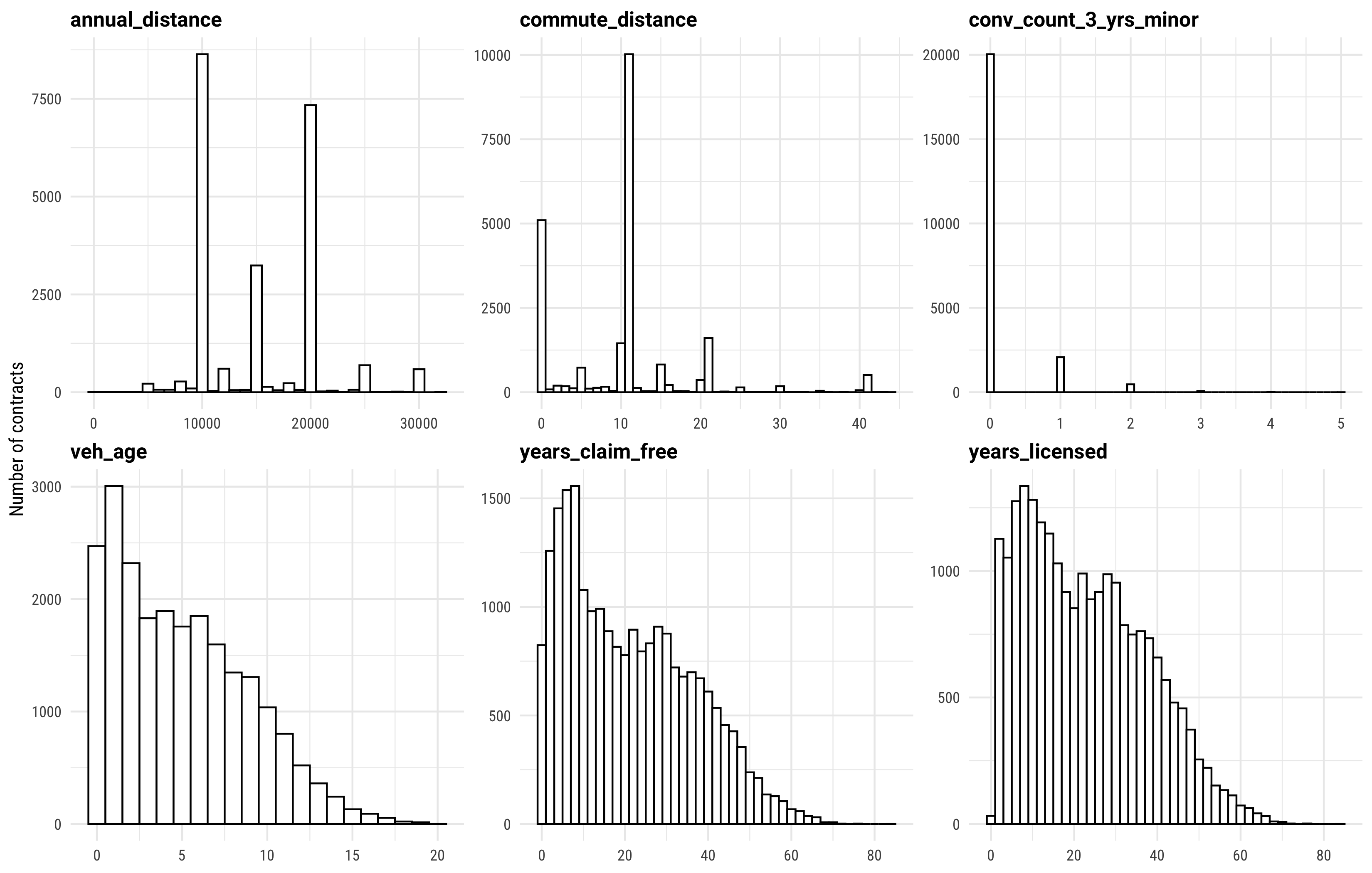}
    \caption{Distributions over the $\numprint{29799}$ vehicles of the 6 classical numeric features selected for the analysis.}
    \label{fig:dist_num}
\end{figure}
\begin{figure}[ht]
    \centering
    \includegraphics[width = \textwidth]{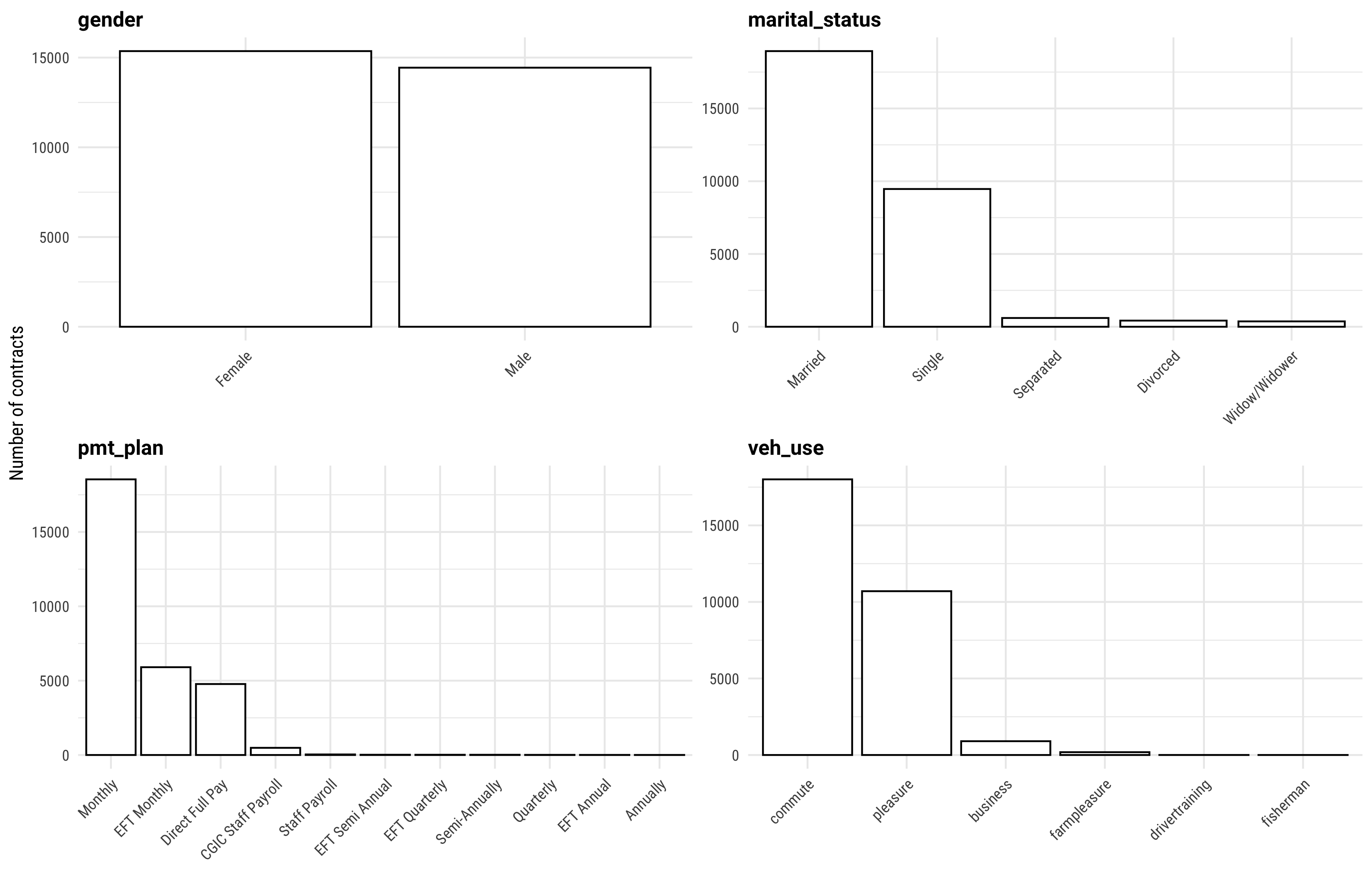}
    \caption{Distributions over the $\numprint{29799}$ vehicles of the 4 classical categorical features selected for the analysis.}
    \label{fig:dist_cat}
\end{figure}

\begin{figure}[ht]
    \centering
    \includegraphics[width = \textwidth]{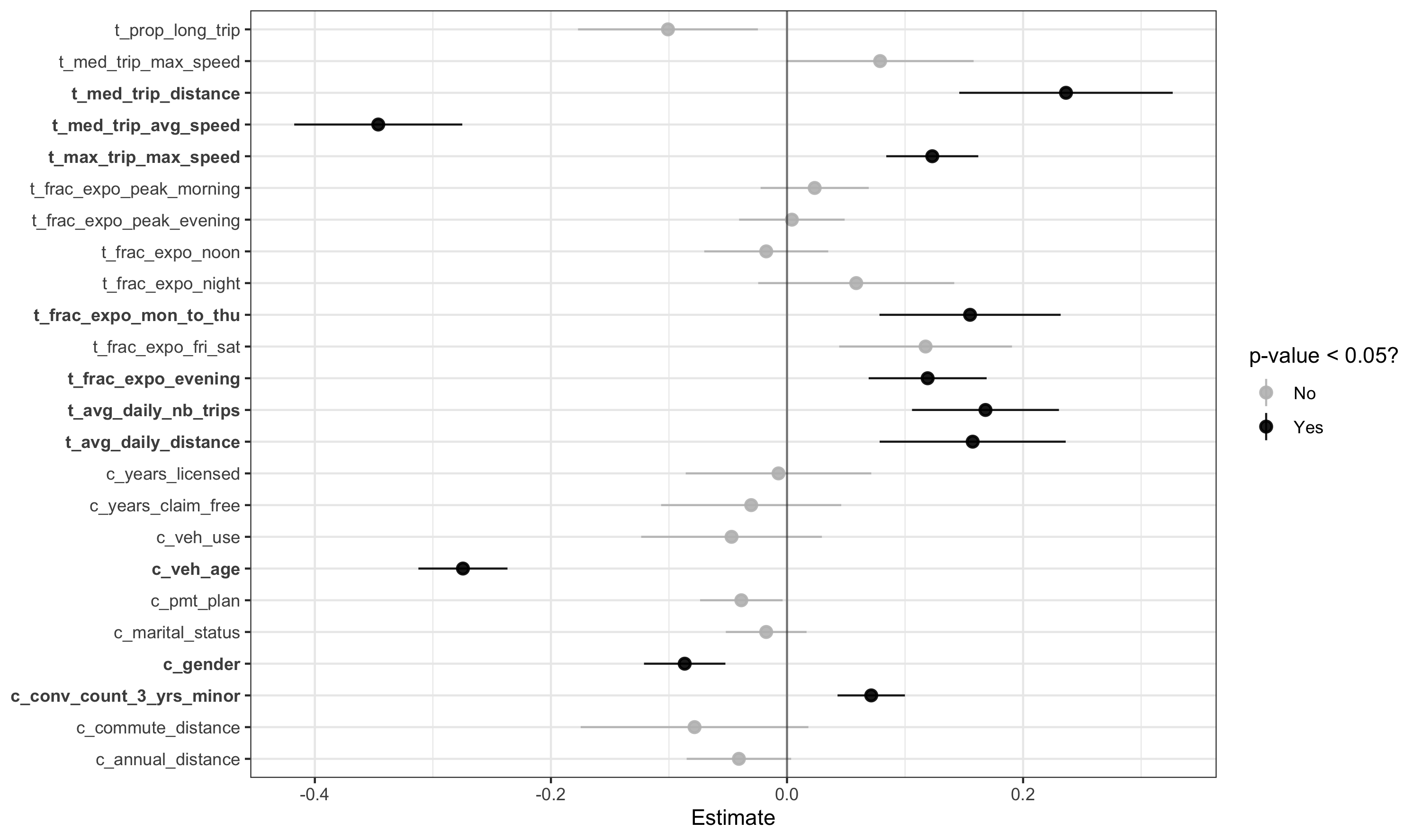}
    \caption{Non-penalized logistic regression coefficients and their standard deviation for classical and telematics features obtained with the $\numprint{29799}$ vehicles.}
    \label{fig:glm_coefs}
\end{figure}


\end{document}